\documentclass[submission, Phys]{SciPost}

\pdfoutput=1
\usepackage{comment}
\usepackage{hyperref}
\usepackage{enumerate}
\usepackage{amssymb}
\usepackage{amsmath}
\usepackage{braket}
\usepackage{graphicx}

\newcommand{\be}{\begin{equation}}
\newcommand{\ee}{\end{equation}}
\newcommand{\ba}{\begin{aligned}}
\newcommand{\ea}{\end{aligned}}
\newcommand{\bw}{\begin{widetext}}
\newcommand{\ew}{\end{widetext}}

\renewcommand{\vec}[1]{\boldsymbol{#1}}
\newcommand{\bea}{\begin{eqnarray}}
\newcommand{\eea}{\end{eqnarray}}

\newcommand{\II}{\mathrm{i}}

\begin{document}

\begin{center}{\Large \textbf{
Entanglement gap, corners, and symmetry breaking
}}\end{center}

\begin{center}
Vincenzo Alba\textsuperscript{1*}
\end{center}

\begin{center}
{\bf 1}	Institute  for  Theoretical  Physics, Universiteit van Amsterdam,
Science Park 904, Postbus 94485, 1098 XH Amsterdam,  The  Netherlands
\\
* v.alba@uva.nl
\end{center}

\begin{center}
\today
\end{center}


\section*{Abstract}
{\bf
We investigate the finite-size scaling of the lowest entanglement gap 
$\delta\xi$ in the ordered phase of the two-dimensional quantum spherical model (QSM). 
The entanglement gap decays as $\delta\xi=\Omega/\sqrt{L\ln(L)}$. This is in contrast with 
the purely logarithmic behaviour as $\delta\xi=\pi^2/\ln(L)$ at the critical 
point. The faster decay in the ordered phase reflects the presence of magnetic 
order. We analytically determine the constant $\Omega$, which depends on the low-energy part 
of the model dispersion and on the geometry of the bipartition. In particular, 
we are able to compute the corner contribution to $\Omega$, at least for the 
case of a square corner. 
}



\section{Introduction}

\begin{figure}[t]
\begin{center}
\includegraphics[width=.5\textwidth]{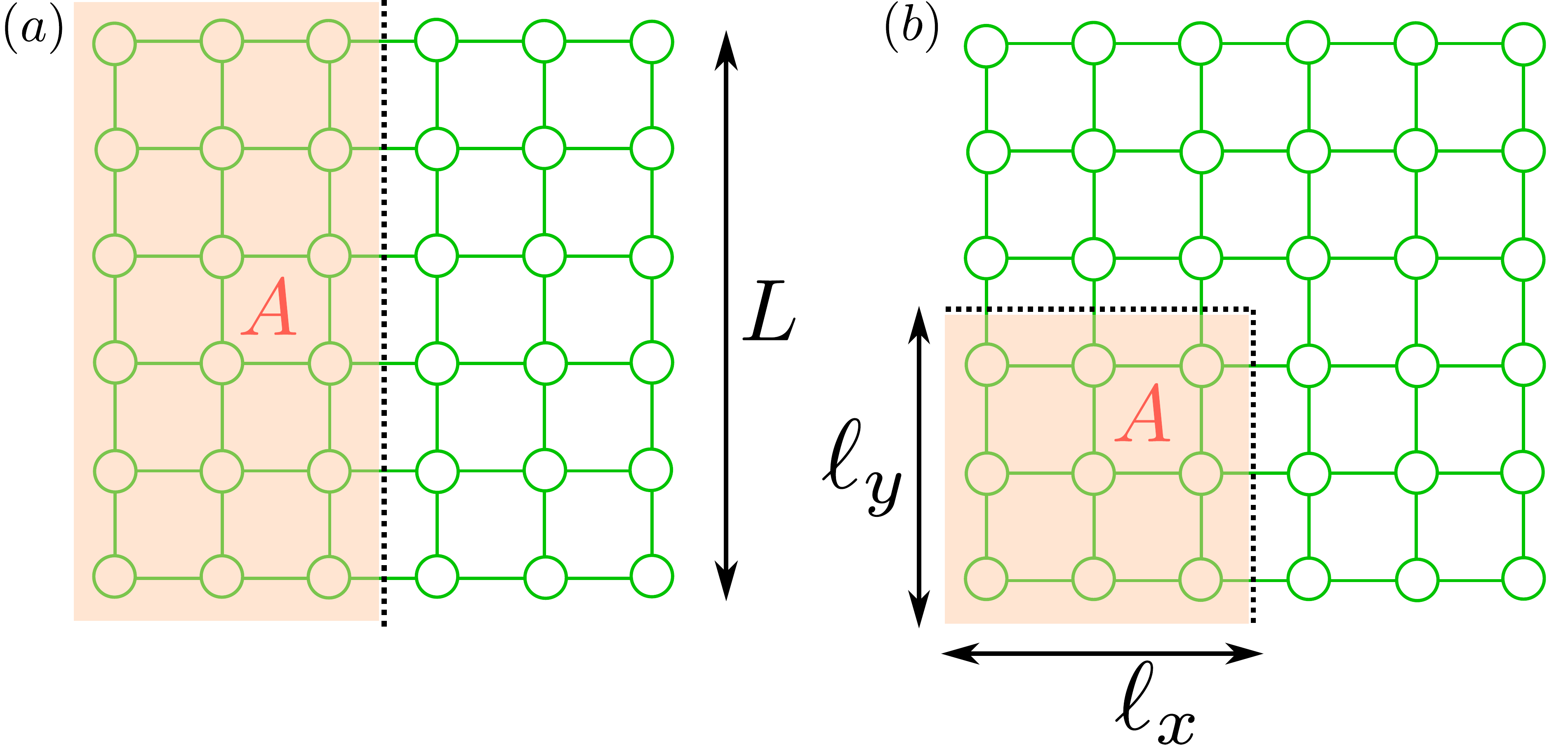}
 \caption{Bipartition of the system as $A\cup\bar A$ used in this work.
 The lattice has $L\times L$ sites and periodic boundary conditions in both 
 directions are used. (a) Bipartition with smooth boundary. 
 (b) Bipartition with a corner. We define the aspect ratios 
 $\omega_{x}=\ell_{x}/L$ and $\omega_y=\ell_y/L$. 
}
\end{center}
\label{fig:partition1}
\end{figure}

In recent years the cross-fertilization between condensed matter and 
quantum information fueled an impressive progress in our understanding 
of quantum many-body systems~\cite{area,amico-2008,calabrese-2009,laflorencie-2016}. 
The entanglement spectrum (ES) has been the subject of intense investigation. 
Let us consider a system in its ground state $|\Psi\rangle$ and a spatial bipartition 
of it as $A\cup\bar A$ (see Fig.~\ref{fig:partition1}). 
The reduced density matrix $\rho_A=\mathrm{Tr}_{\bar{A}}|\Psi\rangle\langle\Psi|$ 
of $A$ can be written as 

\begin{equation}
	\rho_A=e^{-{\mathcal H}_A}. 
\end{equation}
Here ${\mathcal H}_A$ is the so-called entanglement hamiltonian. 
The entanglement spectrum levels $\xi_i=-\ln(\lambda_i)$, with $\lambda_i$ the 
eigenvalues of $\rho_A$, are the ``energies'' of ${\mathcal H}_A$.  
Early works~\cite{chung-2000,peschel-2004,viktor} on 
entanglement spectra aimed at understanding the effectiveness of 
the density matrix renormalisation group (DMRG)~\cite{white-1992,uli-2011} 
to simulate one-dimensional systems. 

Recently, an intense theoretical activity has been devoted to understand the 
ES in fractional quantum Hall systems~\cite{thomale-2010,andreas-2010,haque-2007,thomale-2010a,hermanns-2011,chandran-2011,qi-2012,liu-2012,sterdyniak-2012,dubail-2012,dubail-2012a,Chan14}, topologically ordered systems~\cite{pollmann-2010,turner-2011,bauer-2014}, magnetically 
ordered systems~\cite{poilblanc-2010,cirac-2011,de-chiara-2012,alba-2011,metlitski-2011,alba-2012,Alba13,lepori-2013,james-2013,kolley-2013,Chan14,rademaker-2015,kolley-2015,frerot-2016}, Conformal Field 
Theories (CFTs)~\cite{calabrese-2008,lauchli-2013,Alba-2017,cardy-talk}, and systems with impurities~\cite{bayat-2014}. 
The entanglement gap (or Schmidt gap) $\delta\xi$ emerged as a natural quantity 
to investigate. $\delta\xi$ is the gap of the entanglement hamiltonian, 
and it is defined as 
\begin{equation}
\label{eq:gap-def}
\delta\xi=\xi_1-\xi_0,
\end{equation}
where $\xi_0$ and $\xi_1$ are the first two low-laying ES levels. 
For the standard {\it energy} gap, i.e., the gap of the physical 
hamiltonian, there exists a  ``universal'' correspondence 
between its scaling behaviour and ground state properties, 
such as the decay of correlation functions~\cite{Hastings-2006}. 
Much less is known for the entanglement gap, although several results 
are available. For instance, its behaviour at one-dimensional 
quantum critical points has been 
investigated~\cite{truong-1987,peschel-1999,chung-2000,peschel-2004,alba-2011,andreas-2010,
de-chiara-2012,lepori-2013,di-giulio-2020}. 
In CFTs it is well established that $\delta\xi$ decays logarithmically
as $\delta\xi\propto 1/\ln(\ell)$ with $\ell$ the subsystem length~\cite{calabrese-2008}. 
Similar scaling is found in models that are solvable via 
the corner transfer matrix technique~\cite{truong-1987}. 
Higher-dimensional models are uncharted territory. 
Interestingly, some explicit counterexamples show that the closure of 
the entanglement gap in general does not signal criticality~\cite{Chan14}, 
also for the momentum-space ES~\cite{lundgren-2014}. 
The scenario is different deep in ordered phases of matter. 
For instance, the lower part of the ES of magnetically-ordered 
ground states that break a continuous symmetry~\cite{metlitski-2011} 
is reminiscent of the Anderson 
tower-of-states~\cite{lhullier,beekman-2020,Wietek-2017}. 
This has been verified in systems of quantum rotors~\cite{metlitski-2011}, 
in the two-dimensional Bose-Hubbard model in the 
superfluid phase~\cite{Alba13} (see also \cite{frerot-2016}), 
and also in Heisenberg antiferromagnets on the square~\cite{kolley-2013} and 
on the kagome lattice~\cite{kolley-2015}. 
In the tower-of-states scenario gaps in the lower part of the 
ES decay as a power-law with the subsystem volume, with 
multiplicative logarithmic corrections~\cite{metlitski-2011}. 
Higher ES gaps exhibit a slower 
decay~\cite{metlitski-2011,Alba13,rademaker-2015}. 

Given the lack of general results, exactly solvable models 
can provide valuable insights into the generic features of the entanglement gap. 
Here we investigate the entanglement gap in the ordered phase of the 
two-dimensional quantum spherical model (QSM)~\cite{Ober72,Henk84,Vojta96,Wald15,Bien12}. 
Despite its appealing simplicity, the QSM contains several salient features 
of generic quantum many-body systems. The model is mappable to a system 
of free bosons with an external constraint, implying that its properties can be 
studied with moderate cost. 
Its classical version proved to be valuable to validate the 
theory of critical phenomena and finite size scaling~\cite{brezin-1982}. 
The ground-state phase diagram of the two-dimensional 
QSM exhibits a paramagnetic (disordered)
phase and a ferromagnetic (ordered) one, which are divided  by 
a continuous quantum phase transition. The universality class 
is that of the three-dimensional classical $O(N)$ 
vector model~\cite{zinn-justin-1998} in the large $N$ 
limit~\cite{Stan68,Henk84,Vojta96}.
Entanglement properties of $O(N)$ models have been addressed in 
the past~\cite{Metlitski-2009,Whitsitt-2017} (see 
also~\cite{Lu19,Lu20,Wald20,wald-2020} for recent studies in the QSM). 
Here we consider a two-dimensional lattice of linear size $L$. 
The typical bipartitions that we use are reported in Fig.~\ref{fig:partition1}. 
Figure~\ref{fig:partition1} (a) shows a bipartition with a straight boundary. 
The bipartition in Fig.~\ref{fig:partition1} (b) contains a square corner. 
The effect of corners in the scaling of entanglement-related quantities 
is nontrivial, and it has been studied intensely in the 
last decade~\cite{Casini:2008as,Casini:2006hu,PhysRevLett.110.135702,pitch,2014arXiv1401.3504K,Singh2012,Helmes2014,laflorencie-2016,Seminara-2017}.

Our main result is that in the ordered phase of the QSM, 
in the limit $L,\ell_x,\ell_y\to\infty$ 
with the ratios $\omega_{x,y}=\ell_{x,y}/L$ (see Fig.~\ref{fig:partition1}) 
fixed the entanglement gap decays as 
\begin{equation}
\label{eq:main-res}
\delta\xi=\frac{\Omega}{\sqrt{L\ln(L)}}+\dots 
\end{equation}
Here the dots denote subleading terms that we neglect. The constant 
$\Omega$, which we determine analytically, depends on the low-energy 
properties of the model and on the geometry of the bipartition. 
In particular, we analytically determine the corner contribution to 
$\Omega$. 
The ``fast'', i.e., power-law behaviour as $1/\sqrt{L}$ in~\eqref{eq:main-res} 
reflects the 
presence of magnetic order, whereas the logarithmic correction is 
similar to the critical behaviour~\cite{wald-2020} of $\delta\xi$. 
Finally, we should mention that Eq.~\eqref{eq:main-res} is different 
from the result derived in Ref.~\cite{metlitski-2011}, where it was shown 
that for $O(N)$ models $\delta\xi\propto (L\ln(L))^{-1}$. 
The discrepancy could be explained by the fact that 
the breaking of the $O(N)$ symmetry that is associated with the 
onset of magnetic order happens only in the thermodynamic limit. For finite-size 
systems the symmetry is preserved. The result of Ref.~\cite{metlitski-2011} are 
derived within this scenario. On the other hand, in the QSM the 
spherical symmetry is imposed only on average, 
even for finite systems. 

The manuscript is organised as follows. In section~\ref{sec:model} we 
introduce the QSM. In section~\ref{sec:fs} we review the finite-size 
scaling of the ground-state two-point correlation functions. In 
section~\ref{sec:es-def} we 
briefly overview the calculation of the entanglement gap in the QSM. 
Section~\ref{sec:approx-2} is devoted to the derivation of our main 
results. In section~\ref{sec:num} we provide numerical checks. 
We conclude in section~\ref{sec:concl}. In Appendix~\ref{app:app1} we report 
some technical derivations. 

\section{Quantum Spherical Model}
\label{sec:model}

\noindent
The quantum spherical model~\cite{Henk84, Vojta96,Wald15} (QSM) on a 
two dimensional square lattice of volume $V=L^2$, with $L$ being the lattice linear size,
is defined by the hamiltonian
\begin{equation}
\label{ham}
H = \frac{g}{2}\sum_{\vec{n}} p_{\vec{}}^2 
- J\sum_{\langle \vec{n},\vec{m}\rangle} s_{\vec{n}}s_{\vec{m}} 
+ (\mu+2)\sum_{\vec{n}} s_{\vec{n}}^2. 
\end{equation}
Here, $n=(n_x,n_y)$ denotes a generic lattice site, and 
$\langle n,m\rangle$ a lattice bond joining two nearest-neighbour sites. 
We set $J=1$ in~\eqref{ham}.
The spin $s_i$  and momenta $p_i$ variables satisfy 
standard bosonic commutation relations 
\begin{equation}
	[p_{\vec{n}},p_{\vec{m}}]=[s_{\vec{n}},s_{\vec{m}}]=0,
	\quad[s_{\vec{n}},p_{\vec{m}}]=\II\delta_{\vec{nm}}. 
\end{equation}
Here we refer to the parameter $g$ as the  {\it quantum coupling}. 
Indeed, in the limit $g\to 0$ the model reduces to the classical 
spherical model~\cite{Berl52,Lew52}. 
The spherical parameter $\mu$ is a Lagrange multiplier that 
fixes the global magnetization as  
\begin{equation}
\label{constr-mu}
\sum_{\vec{n}} \langle s_{\vec{n}}^2\rangle=V,  
\end{equation}
To diagonalize the QSM hamiltonian~\eqref{ham}, one can 
exploit its translational invariance.  First, one performs 
a  Fourier transform as 
\begin{equation}
	p_{\vec{n}} = \frac{1}{\sqrt{V}}
	\sum_{\vec{k}} e^{-\II \vec{n} \vec{k}}\pi_{\vec{k}}\ ,
	\qquad s_{\vec{n}} = \frac{1}{\sqrt{V}}
	\sum_k e^{\II \vec{n} \vec{k}}q_{\vec{k}}, 
\end{equation}
where the sum is over $\vec{k}= (k_x,k_y)$ in the first Brillouin zone 
$k_i= 2\pi/L j$, with $j\in[-L/2,L/2]$ integer. In Fourier space one 
obtains 
\begin{equation}
\label{ham-k}
 H = \sum_{\vec{k}} \frac{g}{2}\pi_{\vec{k}} \pi_{-{\vec{k}}}
 + \Lambda_{\vec{k}}^2 \, q_{\vec{k}}q_{-{\vec{k}}}. 
\end{equation}
The single-particle dispersion relation is given as 
\begin{equation}
\label{disp}
	\Lambda_{\vec{k}} = \sqrt{\mu + \omega_{\vec{k}}}\quad \textrm{with} \quad 
	\omega_{\vec{k}} = 2-\cos k_x-\cos k_y. 
\end{equation}
To fully diagonalise~\eqref{ham-k} we introduce the new bosonic ladder 
operators $b_{\vec{k}}$ and $b_{\vec{k}}^\dagger$ as 
\begin{equation}
\label{eq:quantisation}
q_{\vec{k}} = \alpha_{\vec{k}} \frac{b_{\vec{k}}+b_{-\vec{k}}^\dagger}{\sqrt{2}} \ , 
\qquad \pi_{\vec{k}} = \frac{\II}{\alpha_{\vec{k}}}
\frac{b_{\vec{k}}^\dagger - b_{-\vec{k}}}{\sqrt{2}}, 
\end{equation}
where $\alpha_{\vec{k}}^2 = \sqrt{g/2}\Lambda_{\vec{k}}^{-1}$.
Now, the hamiltonian~\eqref{ham-k} is fully diagonal, and it reads as 
\begin{equation}
\label{ham-diag}
H = \sum_{\vec{k}} E_{\vec{k}} (b_{\vec{k}}^\dagger b_{\vec{k}} + 1/2),
\, \textrm{with} \quad E_{\vec{k}} = \sqrt{2g} \Lambda_{\vec{k}}. 
\end{equation}
For the following, it is useful to consider the 
ground-state two-point correlation functions $\langle s_{\vec{n}}
s_{\vec{m}}\rangle$ and $\langle p_{\vec{n}}p_{\vec{m}}\rangle$. 
They are given as~\cite{Wald15} 
\begin{align}
\label{eq:snsm}
\mathbb{S}_{\vec{nm}}=\langle s_{\vec{n}} s_{\vec{m}} \rangle &= \frac{1}{2V}\sum_{\vec{k}} 
e^{\II (\vec{n}-\vec{m})\cdot \vec{k}}  \alpha_{\vec{k}}^2 \\
\label{eq:pnpm}
\mathbb{P}_{\vec{nm}} =\langle p_{\vec{n}} p_{\vec{m}} \rangle &= \frac{1}{2V}\sum_{\vec{k}}
e^{-\II (\vec{n}-\vec{m})\cdot \vec{k}}  \alpha_{\vec{k}}^{-2}\\
\label{eq:snpm}
\mathbb{K}_{\vec{nm}}=\langle s_{\vec{n}} p_{\vec{m}}\rangle &= \frac{\II}{2}\delta_{\vec{nm}}
\end{align}
Importantly, the trivial identity holds 
\begin{equation}
\label{eq:id}
\mathbb{P}_{\vec{n}\vec{m}}=\frac{1}{g}\int d\mu\, \mathbb{S}_{\vec{n}\vec{m}}. 
\end{equation}
By using~\eqref{eq:snsm}, the spherical constraint~\eqref{constr-mu} can be 
rewritten as
\begin{equation}
\label{mu-fs}
\frac{2}{g} = \frac{1}{V}\sum_{\vec{k}}\frac{1}{E_{\vec{k}}}=
\frac{2}{g}\mathbb{S}_{\vec{n}\vec{n}}. 
\end{equation}
Eq.~\eqref{mu-fs} is the so-called gap equation in the context of the 
large-$N$ model~\cite{Amit84}. 
A crucial observation is that the correlator~\eqref{eq:snsm} 
exhibits a singularity for $\vec{k}=0$. This zero mode will play a crucial 
role in the behaviour of the entanglement gap. 

In two dimensions at zero temperature the QSM exhibits a 
second-order phase transition at a critical 
value $g_c$. The value of $g_c$ is known analytically as 
\begin{equation}
  g_c= \frac{\pi^4}{2} K^{-4}\left(1/2-1/\sqrt{2}\right) \simeq 9.67826. 
  \label{gcrit}
\end{equation}
For $g<g_c$ the QSM exhibits a magnetically ordered phase, which is the focus of 
this work. At $g>g_c$ the ground state is paramagnetic. Different phases are 
associated with different behaviours of the spherical parameter $\mu$. 
In the paramagnetic phase one has 
that $\mu$ is  nonzero. On the other hand, $\mu=0$  at the critical point, and in 
the ordered phase. 
The different phases of the model correspond to different finite-size scaling behaviours 
of $\mu$. In the paramagnetic phase one has $\mu={\mathcal O}(1)$. 
At the critical point one can show that $\mu={\mathcal O}(1/L^2)$. 
In the ordered phase $\mu={\mathcal O}(1/L^4)$. 
The critical behaviour at $g_c$ is in the universality class of the three-dimensional 
$N$-vector model~\cite{Vojta96} at large $N$. 

\section{Spin and momentum correlators}
\label{sec:fs}

%
\begin{figure}[t]
\begin{center}
\includegraphics[width=.6\textwidth]{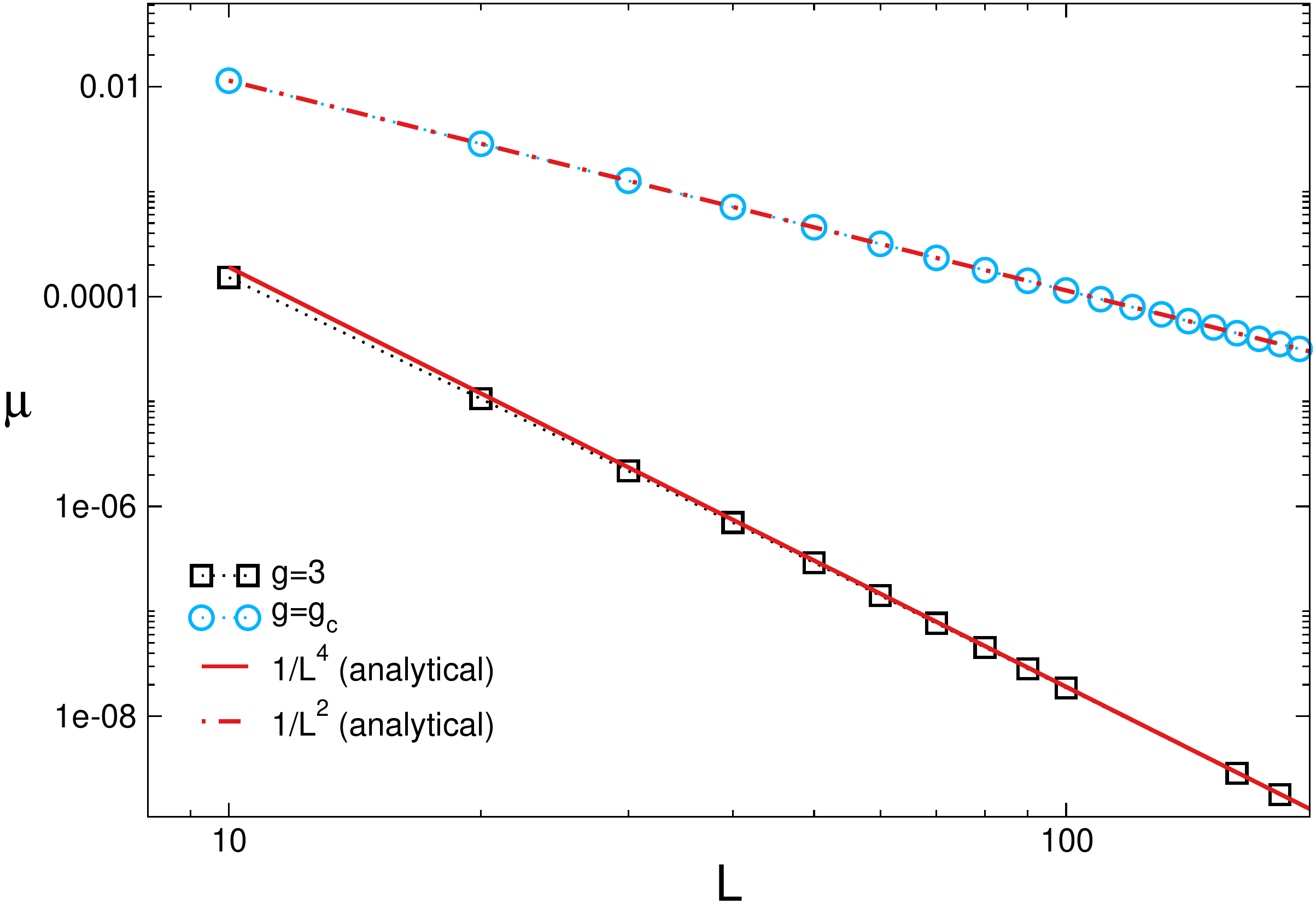}
\caption{ Spherical parameter $\mu$ at the critical point at 
 $g_c\simeq9.67$ and in the ordered phase at $g=3$. Symbols are 
 exact numerical results. The lines are analytical results in 
 the large $L$ limit. 
}
\end{center}
\label{fig:mu}
\end{figure}
%
Here we summarise the finite-size scaling of the spin-spin correlation function 
$\mathbb{S}_{\vec{n}\vec{m}}$ (cf.~\eqref{eq:snsm}) and the momentum correlation 
function $\mathbb{P}_{\vec{n}\vec{m}}$ (cf.~\eqref{eq:pnpm}). 
Let us focus first on the spin correlator. We are interested in the limit 
$L\to\infty$. We can decompose the correlator as 
\begin{equation}
\label{eq:s-decomp}
{\mathbb{S}}_{\vec{n}\vec{m}}=
\mathbb{S}_{\vec{n}\vec{m}}^{(th)}+\mathbb{S}_{\vec{n}\vec{m}}^{(L)}+\dots 
\end{equation}
The first term is the leading term in the large $L$ limit. Note that the first term 
depends on $L$ because $\mu$ depends on $L$. The second term in~\eqref{eq:s-decomp} 
is the first subleading correction in powers of $1/L$. The dots denote more subleading terms 
that we neglect. The thermodynamic contribution is given as  
\begin{equation}
	\label{eq:snm-th}
	\mathbb{S}_{\vec{n}\vec{m}}^{(th)}=
	\frac{\sqrt{g_c}}{2\sqrt{2}(2\pi)^2}\int d\vec{k}\frac{e^{i\vec{k}(\vec{n}-\vec{m})}}
	{\sqrt{\mu+\omega_{\vec{k}}}}, 
\end{equation}
The finite-size part has the surprisingly simple form~\cite{wald-2020} as  
\begin{equation}
\label{eq:s1}
\mathbb{S}^{(L)}_{\vec{n}\vec{m}}=
\frac{\sqrt{g}}{4\pi }\sideset{}{'}
\sum_{l,l'=-\infty}^\infty\frac{
	e^{-\sqrt{2\mu}F_{ll'}({\vec{n}},{\vec{m}})}}
	{F_{ll'}({\vec{n}},{\vec{m}})}. 
\end{equation}
Here we defined 
\begin{equation}
\label{eq:f-def}
F_{ll'}(\vec{n},\vec{m})=\sqrt{(l L+n_x-m_x)^2+(l' L+n_y-m_y)^2}. 
\end{equation}
The prime in the sum means that one has to remove the term 
with $(l,l')=(0,0)$. Eq.~\eqref{eq:s1} holds in the limit 
$L\to\infty$ and $\mu\to0$, i.e., for $g\le g_c$. 
The correlators $\mathbb{S}_{\vec{n}\vec{m}}$ depend only on 
$n_x-m_x$ and $n_y-m_y$, reflecting translation invariance. Moreover, 
the infinite sums in $l,l'$ enforces that $\mathbb{S}_{\vec{n}\vec{m}}$ 
is periodic along the two directions, i.e., it is invariant under 
$n_y-m_y\to n_y-m_y\pm L$ and $n_x-m_x\to n_x-m_x\pm L$. 
Interestingly, $\mathbb{S}_{\vec{n}\vec{m}}^{(L)}$ is singular 
if either $\omega_y=1$  or $\omega_x=1$ (see Fig.~\ref{fig:partition1} (a)), 
whereas no singularity occurs for $\omega_x<1$ and $\omega_y<1$, i.e, in the 
presence of a bipartition with a corner (see Fig.~\ref{fig:partition1} (b)). 
Let us consider the case $\omega_y=1$. Now the terms with $l=0$ and 
$l'=\pm1$ in~\eqref{eq:s1} are singular in the limit $n_x-m_x\to0$ 
and $n_y-m_y\to \pm1$. Terms with $|l'|>1$ or $|l|>1$ in~\eqref{eq:s1} 
are not singular, and do not affect the singularity structure 
of $\mathbb{S}_{\vec{n}\vec{m}}$. These singularities will play an important 
role in section~\ref{sec:approx-2}. 

Similar to~\eqref{eq:s-decomp}, we can decompose the momentum correlator as 
\begin{equation}
\label{eq:p-decomp}
\mathbb{P}_{\vec{n}\vec{m}}=
\mathbb{P}_{\vec{n}\vec{m}}^{(th)}+\mathbb{P}_{\vec{n}\vec{m}}^{(L)}+\dots 
\end{equation}
Here we defined
\begin{equation}
\label{eq:p-th}
	\mathbb{P}_{\vec{n}\vec{m}}^{(th)}=
	\frac{1}{4\sqrt{2 g}\pi^2}\int_{-\pi}^\pi d\vec{k} e^{i\vec{k}(\vec{n}-\vec{m})}
	\sqrt{\mu+\omega_{\vec{k}}}. 
\end{equation}
The finite-size part $\mathbb{P}_{\vec{n}\vec{m}}^{(L)}$ has the same 
structure as~\eqref{eq:s1},  and it reads as 
\begin{equation}
\label{eq:q2}
\mathbb{P}^{(L)}_{\vec{n}\vec{m}}=
-\frac{1}{4\pi\sqrt{g}}\,\,\sideset{}{'}\sum_{l,l'=-\infty}^\infty
e^{-\sqrt{2\mu}F_{ll'}(\vec{n},\vec{m})}
\Big[\frac{1}{F_{ll'}^3(\vec{n},\vec{m})}
+\frac{\sqrt{2\mu}}{F_{ll'}^2(\vec{n},\vec{m})}\Big], 
\end{equation}
with $F_{ll'}(\vec{n},\vec{m})$ as defined in~\eqref{eq:f-def}. 
Eq.~\eqref{eq:p-th} and Eq.~\eqref{eq:q2} are obtained 
from~Eq.~\eqref{eq:snm-th} and Eq.~\eqref{eq:s1} by using~\eqref{eq:id}.
As for Eq.~\eqref{eq:s1}, the finite-size term~\eqref{eq:q2} is singular 
if  subsystem $A$ spans the full lattice in one of the two directions, 
i.e., if $\omega_y=1$ or $\omega_y=1$ (see Fig.~\ref{fig:partition1}). 
For $\omega_y=1$ the singularity occurs  for $l=0$ and 
$l'=\pm1$ in the limit $n_x-m_x\to0$ and 
$n_y-m_y\to\pm1$. Finally, the first term has a stronger 
singularity than the second one.

\subsection{Spherical parameter}
\label{sec:mu}

Let us discuss the finite-size scaling of the spherical constraint 
$\mu$ (cf.~\eqref{mu-fs}) in the ordered phase of the QSM. 
For $g\le g_c$ the spherical parameter vanishes in the thermodynamic limit. 
At the critical point one has the behaviour~\cite{wald-2020} 
$\mu\propto \gamma_2^2/(2L^2)$, 
with $\gamma_2$ a universal constant. To derive the behaviour of $\mu$ in 
the ordered phase we use Eq.~\eqref{eq:snsm} in the 
gap equation~\eqref{mu-fs}. We obtain 
\begin{equation}
\label{eq:gap-eq}
\frac{1}{\sqrt{g}}=\frac{1}{2\sqrt{2}(2\pi)^2}\int \frac{d\vec{k}}{\sqrt{\omega_{\vec{k}}}}
-\frac{\sqrt{\mu}}{2\sqrt{2}\pi}
+\frac{1}{4\pi L}\,
\sideset{}{'}\sum_{l,l'=-\infty}^\infty\frac{e^{-\sqrt{2\mu}L\sqrt{l^2+l'^2}}}{\sqrt{l^2+l'^2}}
\end{equation}
As it clear from the exponent in the last term in~\eqref{eq:gap-eq} the 
scaling as $\mu\propto 1/L^2$ at criticality implies that terms with large $l,l'$ are 
exponentially suppressed. On the other hand, for $\mu\propto1/L^4$ this is not 
the case because the term in the exponent in~\eqref{eq:gap-eq} is ${\mathcal O}(1/L)$. 
First, we anticipate that the second term in~\eqref{eq:gap-eq} is ${\mathcal O}(1/L^2)$,  
and it is subleading. 
To extract the leading behaviour of $\mu$ we use the very elegant identity involving the 
function ${\mathcal K}(\sigma,d,y)$ defined as~\cite{singh-1989}  
\begin{equation}
\label{eq:K-pathria}
{\mathcal K}(\sigma,d,y)=\sideset{}{'}\sum\limits_{\vec{l}(d)}\frac{K_\sigma(2y |\vec{l}|)}{(y|\vec{l}|)^\sigma},\quad 
|\vec{l}|=(l_1^2+l_2^2+\dots+l_d^2)^{\frac{1}{2}}. 
\end{equation}
Here the sum is over the $d$-dimensional vector of integers $l_i\in(-\infty,\infty)$,  
$K_\sigma(z)$ is the modified Bessel function of the second kind, 
and $y>0$ and $\sigma$ are real parameters. We are interested in the case 
$d=2$ and $\sigma=1/2$ (cf.~\eqref{eq:gap-eq}). 
One can show that~\cite{singh-1989}  
\begin{multline}
	\label{eq:beauty-id}
	{\mathcal K}=\frac{1}{2}\pi^\frac{d}{2}\Gamma\Big(\frac{d}{2}-\sigma\Big)y^{-d}
	+\frac{1}{2}\pi^{2\sigma-\frac{d}{2}}C(\sigma,d)y^{-2\sigma}-\frac{1}{2}\Gamma(-\sigma)\\+
	\frac{1}{2}\pi^{2\sigma-\frac{d}{2}}\Gamma\Big(\frac{d}{2}-\sigma\Big)y^{-2\sigma}
	\sideset{}{'}\sum\limits_{\vec{l}(d)}\Big[\Big(|\vec{l}|^2+\frac{y^2}{\pi^2}\Big)^{\sigma
	-\frac{d}{2}}-|\vec{l}|^{2\sigma-d}\Big].
\end{multline}
The constant $C(\sigma,d)$ for $d=2$ reads as 
\begin{equation}
\label{eq:cnu}
C(\sigma,2)=4\Gamma(1-\sigma)\zeta(1-\sigma)\beta(1-\sigma),
\end{equation}
where $\zeta(x)$ is the Riemann zeta function, and 
$\beta(x)$ is the analytic continuation of the Dirichlet series 
\begin{equation}
\label{eq:beta}
\beta(x)=\sum_{l=0}^\infty\frac{(-1)^l}{(2l+1)^x}. 
\end{equation}
To apply~\eqref{eq:beauty-id} we fix $y=\sqrt{\mu/2}L$. 
In the limit $\mu\to0$ the leading behaviour of ${\mathcal K}$ 
is given by the first term on the right hand side in~\eqref{eq:beauty-id}. 
After using that in~\eqref{eq:gap-eq} we obtain
\begin{equation}
\label{eq:gap-1}
\frac{1}{\sqrt{g}}=\frac{1}{8\pi^2\sqrt{2}}\int\frac{d\vec{k}}{\sqrt{\omega_{\vec{k}}}}
+\frac{1}{2\sqrt{2\mu}L^2}. 
\end{equation}
In~\eqref{eq:gap-1} we are neglecting vanishing terms in the limit $L\to\infty$. 
The second term in~\eqref{eq:gap-1} is also simply obtained by isolating the term with 
$\vec{k}=0$, i.e., the zero mode, in the sum in~\eqref{mu-fs}. 
It is now clear that we can parametrize $\mu$ as 
\begin{equation}
\label{eq:mu-redef}
\mu=\frac{\gamma_4^2}{L^4}.
\end{equation}
After substituting in~\eqref{eq:gap-1}, we obtain that 
\begin{equation}
\label{eq:gamma4}
\gamma_4=\Big[
	\frac{2\sqrt{2}}{\sqrt{g}}-\frac{1}{4\pi^2}\int\frac{d\vec{k}}{\sqrt{\omega_{\vec{k}}}}
\Big]^{-1}
\end{equation}
Note that the constant $\gamma_4$ is not universal, as it is clear from the 
explicit dependence on $g$. This is expected, and it is in contrast with the 
result at the critical point, where $\mu=\gamma_2^2/(2L^2)$, with $\gamma_2$ 
universal. 

\section{Entanglement gap in the QSM}
\label{sec:es-def}

Here we briefly review how to calculate the entanglement gap in the 
QSM. Entanglement properties of the QSM are derived from the two-point 
correlation functions~\eqref{eq:snsm} and~\eqref{eq:pnpm} 
because the model can be mapped to free bosons (see Ref.~\cite{viktor} for a review). 
We first define the correlation matrix $\mathbb{C}$ restricted to subsystem $A$ as 
\begin{equation}
\label{eq:ca}
\mathbb{C}_A=\mathbb{S}_A \cdot \mathbb{P}_A, 
\end{equation}
with $\mathbb{S}_A$ and $\mathbb{P}_A$ defined in~\eqref{eq:snsm} 
and~\eqref{eq:pnpm}, with $\vec{n},\vec{m}\in A$. 
Since the QSM is mapped to free-bosons, the reduced density matrix of a 
subsystem $A$ is a quadratic operator, and it is written as~\cite{viktor}  
\begin{equation}
	\label{eq:rdm-fb}
	\rho_A=Z^{-1}e^{-{\mathcal H}_A},\quad{\mathcal H}_A=\sum_k\epsilon_k b^\dagger_k 
	b_k. 
\end{equation}
Here ${\mathcal H}_A$ is the so-called entanglement hamiltonian, $\epsilon_k$ are 
{\it single-particle} entanglement spectrum levels, and 
$b_k$ are free-bosonic operators. $Z$ is a normalization factor. 
The eigenvalues $e_k$ of $\mathbb{C}_A$ are obtained from 
the $\epsilon_k$ as  
\begin{equation}
\label{eq:spectra}
\sqrt{e_k} =\frac{1}{2}\coth\left(\frac{\epsilon_k}{2}\right). 
\end{equation}
The entanglement spectrum, i.e., the spectrum of ${\mathcal H}_A$ 
is obtained by filling in all the possible 
ways the single-particle levels $\epsilon_k$. The lowest ES level is the 
vacuum state. Thus, the lowest entanglement gap $\delta\xi$ 
(Schmidt gap) is 
\begin{equation}
\label{eq:dxi-def}
\delta\xi=\epsilon_1,
\end{equation}
with $\epsilon_1$ the smallest single-particle ES level, or equivalently, the 
largest $e_1$ (cf.~Eq.~\eqref{eq:spectra}).

\section{Scaling of the entanglement gap in the ordered phase of the QSM}
\label{sec:approx-2}

In this section we investigate the scaling of the entanglement gap for 
$g<g_c$, i.e., in the ordered phase of the QSM. First, it has been 
numerically observed in Ref.~\cite{wald-2020} that for $g<g_c$, in the 
limit $L\to\infty$ the flat vector $|\vec{1}\rangle$ defined as 
\begin{equation}
\label{eq:flat}
|\vec{1}\rangle=\frac{1}{\sqrt{|A|}}(1,1,\dots,1), 
\end{equation}
with $|A|=\ell_x\ell_y$, is the right eigenvector of $\mathbb{C}_A$ 
corresponding to the largest eigenvalue $e_1$, i.e., the zero-mode 
eigenvector. Moreover, $|\vec{1}\rangle$  is also eigenvector of the 
matrix $\mathbb{S}_A$. It is interesting to investigate the structure 
of the associated eigenvalue. 
This is calculated as 
\begin{equation}
\label{eq:ss}
\langle\vec{1}|\mathbb{S}|\vec{1}\rangle=\frac{1}{|A|}\sum\limits_{\vec{n},\vec{m}\in A}\mathbb{S}_{\vec{n}\vec{m}}. 
\end{equation}
After using~\eqref{eq:s-decomp}, it is straightforward to numerically 
check that the thermodynamic part of the 
correlator $\mathbb{S}^{\scriptscriptstyle(th)}$ for large $L$ gives a subleading  term as 
$L\ln(L)$ in~\eqref{eq:ss} (see section~\ref{sec:num}). 
The leading contribution is given by the finite-size 
part of the correlator $\mathbb{S}^{\scriptscriptstyle(L)}$, and it is  
${\mathcal O}(L^2)$. An important observation is that due to the scaling as 
$\mu=\gamma_4^2/L^4$, the dependence on 
the coordinates $\vec{n},\vec{m}$ in~\eqref{eq:s1} can be neglected. 
Thus, a straightforward calculation yields 
\begin{equation}
	\label{eq:s-exp}
\langle\vec{1}|\mathbb{S}|\vec{1}\rangle=
	\frac{\sqrt{g}\omega_x\omega_y L^2}{2\sqrt{2}\gamma_4}. 
\end{equation}
One should observe that Eq.~\eqref{eq:s-exp} is exactly the 
contribution of $\vec{k}=\vec{0}$ in the sum in~\eqref{eq:snsm}. Physically, this 
means that in the ordered phase of the QSM for $g<g_c$ the leading behaviour of 
the eigenvalue of $\mathbb{S}_A$ associated with the flat vector is simply obtained by 
isolating the term with $\vec{k}=\vec{0}$ in~\eqref{eq:snsm}. This happens because 
of the ``fast'' decay as $\mu\propto 1/L^4$. This is not the case at the critical 
point~\cite{wald-2020}, where $\mu\propto 1/L^2$. 
Moreover, this result suggests that one can decompose the correlator $\mathbb{S}$ as 
\begin{equation}
\label{eq:zero-mode}
\mathbb{S}=s_0|\vec{1}\rangle\langle\vec{1}|+\dots, \quad\textrm{with}\,
s_0=\langle\vec{1}|\mathbb{S}|\vec{1}\rangle. 
\end{equation}
Here $s_0\propto L^2$, and the dots are subleading terms that we neglect. 
By using~\eqref{eq:zero-mode} and the fact that $\mathbb{P}$ is finite 
in the limit $L\to\infty$, it is straightforward to show that 
the eigenvalue $e_1$ of $\mathbb{C}_A=\mathbb{P}_A\cdot\mathbb{S}_A$ 
in the limit $L\to\infty$ is given as (see~\cite{Botero04} and~\cite{wald-2020}) 
\begin{equation}
\label{eq:e1-exp}
	e_1=\langle\vec{1}|\mathbb{S}|\vec{1}\rangle\langle\vec{1}|\mathbb{P}|\vec{1}\rangle. 
\end{equation}
Here we have 
\begin{equation}
\label{eq:p-ex}
\langle\vec{1}|\mathbb{P}|\vec{1}\rangle=\frac{1}{|A|}
\sum\limits_{\vec{n},\vec{m}\in A}\mathbb{P}_{\vec{n}\vec{m}}. 
\end{equation}
To proceed we now show that the expectation value $\langle\vec{1}|\mathbb{P}
|\vec{1}\rangle$ decays as $\ln(L)/L$, i.e.,  with a multiplicative logarithmic correction. 
Note that the same scaling behaviour is observed at the critical point~\cite{wald-2020}. 
The derivation requires minimal modifications as compared with the critical case, 
and it is reported in Appendix~\ref{app:app1}. The main ingredients are 
standard tools in the finite-size scaling theory, such as Poisson's summation 
formula and the Euler-Maclaurin formula. 

Let us discuss the final result. Clearly, we can treat the contribution of the 
thermodynamic part (cf.~\eqref{eq:p-th}) and the finite-size part 
(cf.~\eqref{eq:q2}) separately. Similar to what happens at the critical 
point~\cite{wald-2020}, the finite-size part contributes only if the boundary 
between the two subsystems is straight. 
For simplicity we consider the bipartition with $\omega_x=1/p$ and 
$\omega_y=1/q$, with $p,q\in\mathbb{N}$. Note that for 
$\omega_y<1$ the boundary between the two subsystems is not 
straight, i.e., it has a square corner. One obtains the 
generic thermodynamic contribution as 
\begin{multline}
\label{eq:p-th-1}
\langle\vec{1}|\mathbb{P}^{(th)}|\vec{1}\rangle=\\
\sum_{p'=0}^{p-1}\sum_{q'=0}^{q-1}\int_0^{1/p}d k_x\int_0^{1/q}d k_y
\sin^2(\pi(k_x+p'/p))\sin^2(\pi(k_y+q'/q))\eta_{p',q'}(k_x,k_y). 
\end{multline}
The function $\eta_{p',q'}(k_x,k_y)$ reads as 
\begin{multline}
\label{eq:eta-f}
	\eta_{p',q'}(k_x,k_y)=
	\frac{4}{\pi^3\sqrt{g}}\Big[
\frac{q}{(k_x+p'/p)^2}
	+\frac{p}{(k_y+q'/q)^2}
+p\psi'(1+k_y+q'/q)\\
+\frac{q}{1+k_x+p'/p}
+\frac{q}{2(1+k_x+p'/p)^2}
\frac{q}{6(1+k_x+p'/p)^3}
+\dots\Big]\frac{\ln(L)}{L}. 
\end{multline}
The dots in the brackets denote terms with higher 
powers of $1/(k_x+p'/p)$. These can be derived systematically 
by using the Euler-Maclaurin formula. 
The function $\psi'(x)$ is the first derivative of the digamma function 
$\psi(x)$ with respect to $x$. The behaviour as $\ln(L)/L$ is clearly visible 
in~\eqref{eq:eta-f}.  
Similar to the critical point~\cite{wald-2020},  
$\eta_{p',q'}$ is determined by the low-energy part of the dispersion of 
the QSM. 
Finally, let us consider the finite-size contribution~\eqref{eq:q2}. 
From~\eqref{eq:q2} it is clear that the finite-size correlator is regular for $\omega_y<1$ 
and $\omega_x<1$, whereas it exhibits a singularity for $\omega_y=1$ or 
$\omega_x=1$, i.e., for the case of straight boundary (see Fig~\ref{fig:partition1} (b)). 
For the straight boundary this gives a contribution as $\ln(L)/L$, whereas it can be 
neglected if a corner is present. 
Again, this is exactly the same at the critical point~\cite{wald-2020}. 
The derivation of the singular contribution, which is present only for straight boundary, 
is reported in~\ref{app:app2}. The final result reads 
\begin{equation}
	\label{eq:p-fs}
	\langle\vec{1}|\mathbb{P}^{(L)}|\vec{1}\rangle=-\frac{1}{\sqrt{g}\pi}
	\frac{\ln(L)}{L}. 
\end{equation}
The minus sign in~\eqref{eq:p-fs} implies that the presence of corners 
increases the prefactor of the logarithmic growth of $e_1$. 
After putting together Eq.~\eqref{eq:e1-exp}, Eq.~\eqref{eq:s-exp}, Eq.~\eqref{eq:p-th-1} and Eq.~\eqref{eq:p-fs}, one obtains that 
\begin{equation}
\label{eq:e1-fin}
e_1=\Omega' L\ln(L), 
\end{equation}
where the constant $\Omega'$ encodes information about the geometry of the bipartition 
and the model dispersion. 
In Eq.~\eqref{eq:e1-fin} we neglect subleading terms in the limit $L\to\infty$. 
From~\eqref{eq:e1-fin}, after using~\eqref{eq:spectra} 
one obtains that 
\begin{equation}
\label{eq:dxi-fin}
\delta\xi=\frac{\Omega}{\sqrt{L\ln(L)}}, 
\quad\mathrm{with}\,\, \Omega=\frac{1}{\sqrt{\Omega'}}.  
\end{equation}
%
%
\begin{figure}[t]
\begin{center}
\includegraphics[width=.6\textwidth]{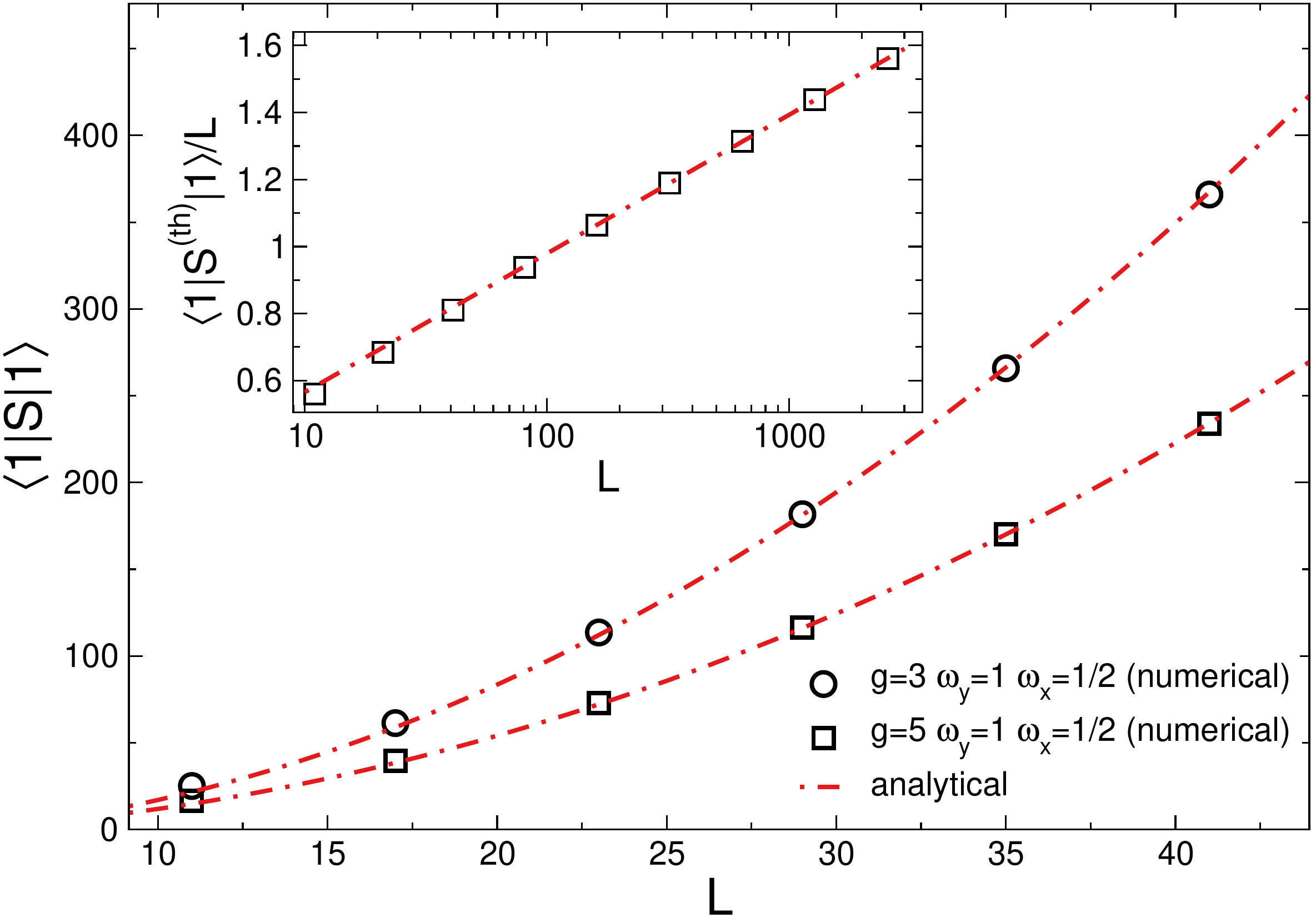}
\caption{ Flat-vector expectation value $\langle\vec{1}|\mathbb{S}
 |\vec{1}\rangle$ of the spin-spin correlator in the ordered 
 phase of the QSM. The behaviour as 
 $\langle\vec{1}|\mathbb{S}|\vec{1}\rangle\propto L^2$ is clearly 
 visible. The dashed-dotted lines are the theory 
 predictions~\eqref{eq:s-exp}. We show results for different 
 aspect ratios $\omega_y,\omega_x$ (see Fig.~\ref{fig:partition1}) 
 and quantum coupling $g$. The inset show the contribution of the 
 thermodynamic part of the correlator $\mathbb{S}^{\scriptscriptstyle (th)}$ 
 (cf.~\eqref{eq:s-decomp}) for $g=5$. The behaviour as $\langle\vec{1}|\mathbb{S}
 ^{\scriptscriptstyle(th)}|\vec{1}\rangle\propto L\ln(L)$ is 
 clearly visible. 
}
\end{center}
\label{fig:s0}
\end{figure}
%
Few comments are in order. First, 
in the ordered phase $\delta\xi$ vanishes in the thermodynamic limit 
as a power law with $L$, except for a logarithmic correction. This is 
different at the critical point, where~\cite{wald-2020} 
$\delta\xi=\pi^2/\ln(L)$. The power-law decay of the entanglement gap in 
symmetry-broken phases has been also numerically observed in magnetic spin systems~\cite{kolley-2013,kolley-2015} and in the ordered phase of the two-dimensional Bose Hubbard 
model~\cite{Alba13}. Note, however, that even with state-of-the-art 
numerical methods it is challenging to observe the logarithmic correction. 
Finally, in Ref.~\cite{metlitski-2011} it has been suggested that in the 
presence of continuous symmetry breaking the gaps in the lower part of the 
entanglement spectrum are 
\begin{equation}
\label{eq:metlitski}
\delta\xi\propto(L\ln(L))^{-1}. 
\end{equation}
This is different from~\eqref{eq:dxi-fin} (note the square root in~\eqref{eq:dxi-fin}). 
The unexpected square root in Eq.~\eqref{eq:dxi-fin} could be explained by the   
way in which in the QSM the spherical constraint is enforced (cf.~\eqref{mu-fs}).  
Further study would be needed to clarify this issue. 
Finally, it is interesting to understand the behaviour of $\delta\xi$ as the critical 
point is approached from the ordered side of the transition. A natural 
scenario is that upon approaching the transition the $1/\sqrt{L}$ is 
``gapped'' out and it gives an extra $1/\sqrt{\ln(L)}$, which allows to 
recover the expected result~\cite{wald-2020} $\delta\xi\propto 1/\ln(L)$.

\section{Numerical results}
\label{sec:num}

In this section we provide numerical evidence supporting the analytic 
result derived in section~\ref{sec:approx-2}. 
%
\begin{figure}[t]
\begin{center}
\includegraphics[width=.6\textwidth]{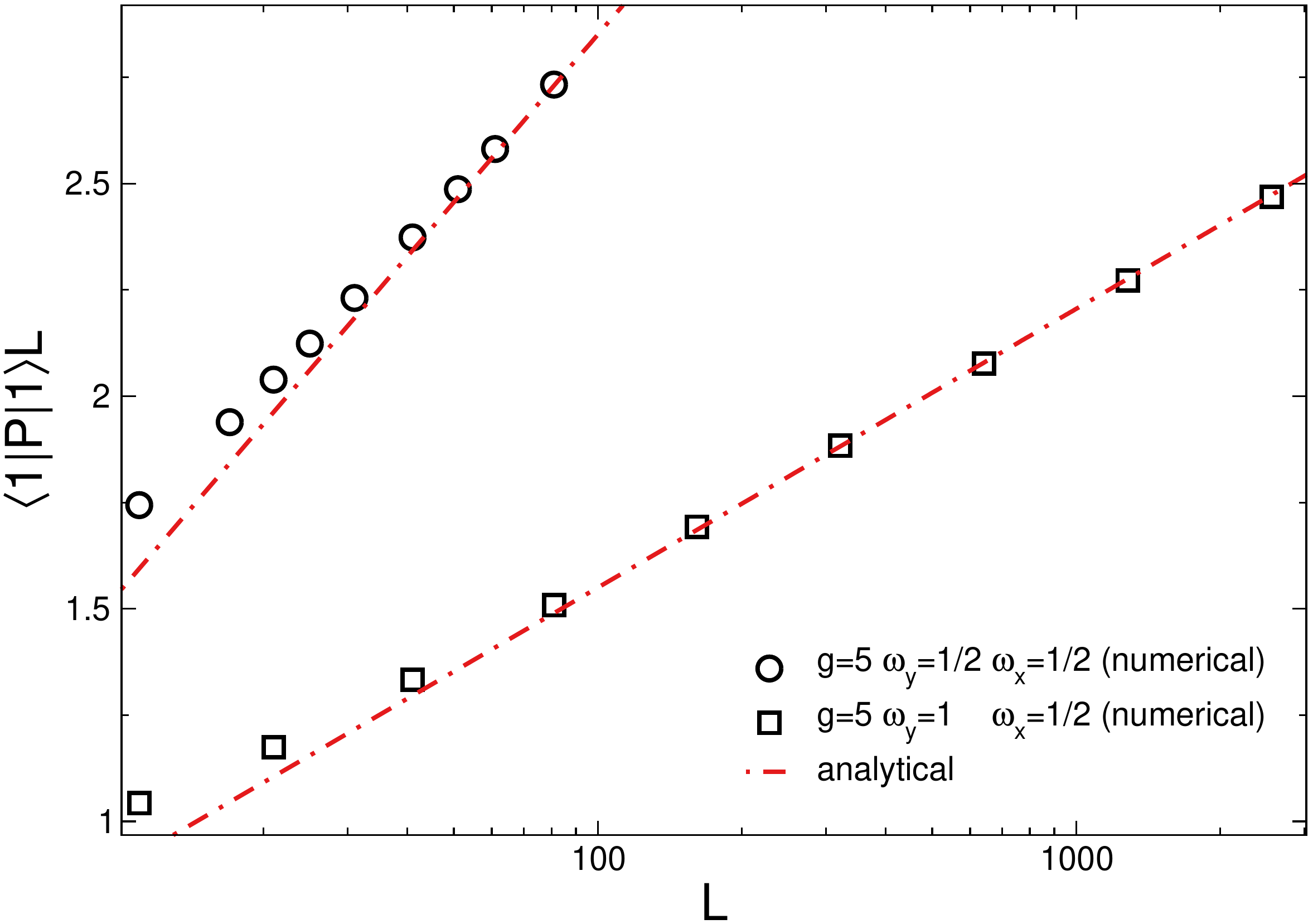}
\caption{Rescaled flat-vector expectation value $\langle\vec{1}|\mathbb{P}
 |\vec{1}\rangle L$ of the momentum operator in the ordered phase of the QSM. 
 We show data for several bipartitions with aspect ratios $\omega_x,\omega_y$ 
 (see Fig.~\ref{fig:partition1}). For $\omega_y<1$ the boundary between 
 the two subsystems is not smooth (see Fig.~\ref{fig:partition1} (b)). Symbols 
 are exact numerical results. The dashed-dotted lines are analytic predictions 
 from~\eqref{eq:p-th-1} and~\eqref{eq:p-fs}. 
}
\end{center}
\label{fig:p0}
\end{figure}
%
Let us start discussing the finite-size scaling of the expectation 
value $\langle\vec{1}|\mathbb{S}|\vec{1}\rangle$. We report numerical 
data in Fig.~\ref{fig:s0}, for fixed $g=3$ (circles) and $g=5$ (squares). 
We only show data for the bipartition with straight boundary $\omega_y=1$ 
(see Figure~\ref{fig:partition1} (a)) and for $\omega_x=1/2$. 
The expected behaviour as $\langle\vec{1}|\mathbb{S}|\vec{1}\rangle\propto L^2$ 
is visible. The dashed-dotted line in the figure is the analytic result 
in Eq.~\eqref{eq:s-exp}, which is in perfect agreement with the numerical 
data. Again, we should stress that Eq.~\eqref{eq:s-exp} originates only 
from the finite-size part $\mathbb{S}^{\scriptscriptstyle(L)}$ (cf.~\eqref{eq:s-decomp}). 
However, it is interesting to investigate the finite-size scaling of the 
flat-vector expectation value calculated using the thermodynamic contribution 
$\mathbb{S}^{\scriptscriptstyle(th)}$. We report this analysis in the 
inset of Fig.~\ref{fig:s0} plotting $\langle\vec{1}|\mathbb{S}^{\scriptscriptstyle(th)}
|\vec{1}\rangle/L$ versus $L$. Data are for $g=5$. Interestingly, the figure shows 
that $\langle\vec{1}|\mathbb{S}^{\scriptscriptstyle(th)}|\vec{1}\rangle\propto L\ln(L)$. 
This confirms that at the leading order in $L$ the expectation value 
$\langle\vec{1}|\mathbb{S}|\vec{1}\rangle$ is dominated by the contribution of the 
zero mode. Finally, we should mention that it 
would be interesting to clarify the origin of the logarithmic 
divergence of the thermodynamic contribution. 

Let us now discuss the flat-vector expectation value of the momentum 
correlator $\langle\vec{1}|\mathbb{P}|\vec{1}\rangle$. In contrast with 
the spin correlator, both the thermodynamic and the finite-size part 
(cf.~\eqref{eq:p-decomp}) contribute to the leading behaviour at large 
$L$. Our numerical data are reported in Fig.~\ref{fig:p0}. In the 
figure we plot $\langle\vec{1}|\mathbb{P}|\vec{1}\rangle L$ versus 
$L$. We show data for $\omega_x=1/2$, $\omega_y=1$ and $\omega_y=1/2$. 
Note that for $\omega_y=1$ the boundary between $A$ and its complement is 
straight. The numerical data in Fig.~\ref{fig:p0} confirm the 
expected behaviour as $\ln(L)/L$ in Eq.~\eqref{eq:p-th-1} and Eq.~\eqref{eq:p-fs}. For 
$\omega_y=1$ the prefactor of the logarithm is obtained by summing Eq.~\eqref{eq:p-th-1} and 
Eq.~\eqref{eq:p-fs}, whereas in the presence of a square corner only Eq.~\eqref{eq:p-th-1} 
has to be considered. 
%
\begin{figure}[t]
\begin{center}
\includegraphics[width=.6\textwidth]{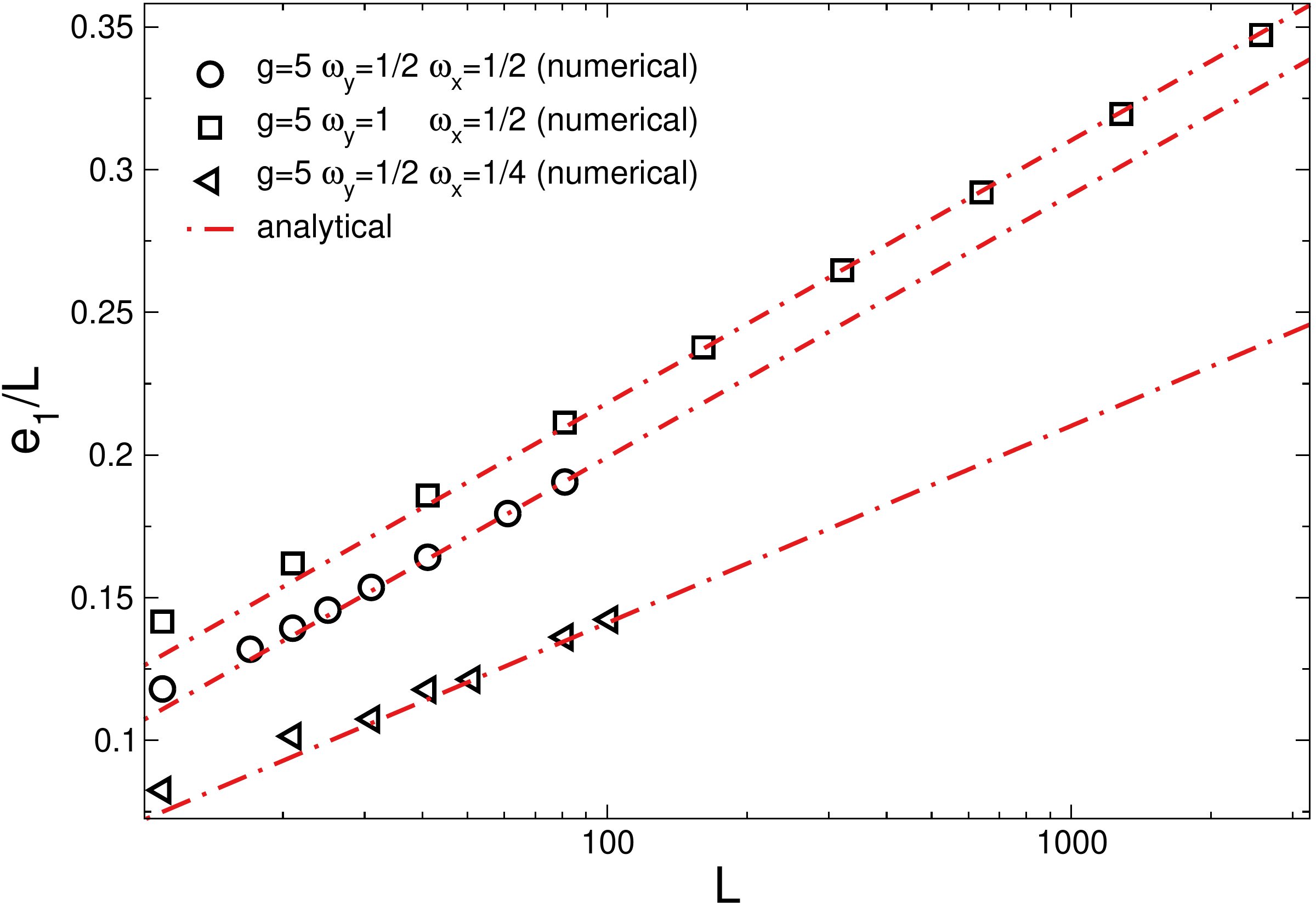}
\caption{ Largest eigenvalue $e_1$ of the correlation matrix $\mathbb{C}$ 
 restricted to $A$. We plot $e_1/L$ versus $L$. Note the logarithmic scale 
 on the $x$-axis. Symbols are exact numerical data. The dashed-dotted lines 
 are analytic predictions. Note that for $\omega_y$ the boundary between 
 the two subsystems has a corner (see Fig.~\ref{fig:partition1} (b)). 
}
\end{center}
\label{fig:eig}
\end{figure}
%
Finally, we discuss the largest eigenvalue $e_1$ of the restricted correlation 
matrix $\mathbb{C}_A$ (cf.~\eqref{eq:ca}). The entanglement gap $\delta\xi$ is 
obtained from $e_1$ via Eq.~\eqref{eq:spectra}. Note that the vanishing of 
$\delta\xi$ is reflected in a diverging $e_1$. We show numerical data for $e_1/L$ 
in Fig.~\ref{fig:eig} plotted versus $L$. We consider several 
aspect ratios $\omega_x$ and $\omega_y$, focusing on $g=5$. In all the 
cases the data exhibit the behaviour $e_1=\Omega' L\ln(L)$. The constant $\Omega'$, which 
depends on the geometry and on low-energy properties of the QSM is obtained by 
combining Eq.~\eqref{eq:e1-exp}  with 
Eq.~\eqref{eq:s-exp}~\eqref{eq:p-th-1}~\eqref{eq:p-fs}. The analytic predictions 
are reported in Fig.~\ref{fig:eig} as dashed-dotted lines and are in perfect 
agreement with the numerical data. This implies that the entanglement gap $\delta\xi$ 
satisfies~\eqref{eq:dxi-fin}.

\section{Conclusions}
\label{sec:concl}

We investigated the entanglement gap in the magnetically ordered phase of the  
two-dimensional QSM. Our main result is that the entanglement gap decays as 
$\delta\xi=\Omega/\sqrt{L\ln(L)}$. We analytically determined the constant 
$\Omega$, which depends on the geometry of the bipartition and on the 
low-energy physics of the model.

There are several intriguing directions for future work. First, it would be interesting 
to explore whether is possible to extend our results to the $N$-vector model at 
finite $N$. An interesting question is whether the discrepancy with the results 
of Ref.~\cite{metlitski-2011} can be attributed to the large $N$ limit. Furthermore, 
it is important to understand how the scaling of the entanglement gap depends on 
dimensionality. This issue could be easily addressed because the QSM is exactly 
solvable in any dimension. Another intriguing direction is to further investigate the role 
of corners. For instance, it would be interesting to investigate the dependence of 
the entanglement gap on the corner angle. It would be also interesting to investigate 
how the outlined scenario is affected by long-range interactions. This should be 
straightforward because the QSM is exactly solvable also in the presence of 
long-range interactions. An exciting possibility is to investigate what happens 
to the entanglement gap in the presence of 
disorder~\cite{Kosterlitz-1976,Jagannathan-1989,Hornreich-1982,Vojta-1993}. 
Finally, a very interesting direction is to study $\delta\xi$ 
after a quantum quench. This could be addressed using 
the results of Ref.~\cite{Barbier-2020}. 

\section*{Acknowledgements}
\label{sec:ack}
I would like to thank Sascha Wald and Raul Arias for several discussions 
in a related project. I acknowledge  support  from  the  European  
Research  Council  under  ERC  Advanced  grant 743032 DYNAMINT.

\appendix

\section{Derivation of the expectation value  
	$\langle\vec{1}|\mathbb{P}|\vec{1}\rangle$}
\label{app:app1}

In this appendix we derive the large $L$ behaviour of the 
expectation value of the momentum correlator with the flat vector $\langle\vec{1}|\mathbb{P}
|\vec{1}\rangle$ (cf.~\eqref{eq:p-ex}). We consider the leading, i.e, the thermodynamic 
limit, as well as the first subleading contributions. The main goal 
is to show that the expectation value exhibits a multiplicative logarithmic 
correction. Two types of contributions are present. One 
originates from the thermodynamic limit of the correlator, 
whereas the second one is due to the first subleading. The latter 
is present only for straight boundary between the two subsystems 
(see Fig.~\ref{fig:partition1}).

\subsection{Thermodynamic contribution}
\label{sec:p-th-c}

Here derive the thermodynamic contribution, which is given 
as $\langle\vec{1}|\mathbb{P}^{(th)}|\vec{1}\rangle$. 
Here $|\vec{1}\rangle$ is the flat vector in region $A$, i.e, 
\begin{equation}
	|\vec{1}\rangle=\frac{1}{\sqrt{|A|}}(1,1,\dots,1),\quad |A|=\ell_x\ell_y. 
\end{equation}
The momentum correlation reads 
\begin{equation}
\label{eq:app-p-th}
	\mathbb{P}_{\vec{n}\vec{m}}^{(th)}=
	\frac{1}{4\sqrt{2g}\pi^2}\int_{-\pi}^\pi d\vec{k} e^{i\vec{k}(\vec{n}-\vec{m})}
	\sqrt{\mu+\omega_{\vec{k}}}. 
\end{equation}
After performing the sum over $\vec{n}$ and $\vec{m}$ in~\eqref{eq:app-p-th}, and after 
changing variables to $k_x'=L\omega_x k_x/\pi$  and $k'_y=L\omega_y k_y/\pi$,  
we obtain  
\begin{multline}
\label{eq:app-p3}
\langle \vec{1}|\mathbb{P}^{(th)}|\vec{1}\rangle=
\frac{2\sqrt{2}}{\sqrt{g}L^4\omega_x^2\omega_y^2}
\int_{0}^{L\omega_x/2} dk_x\int_{0}^{L\omega_y/2} dk_y
\frac{\sin^2(\pi k_x)\sin^2(\pi k_y)}{\sin^2\big(\frac{\pi}{L\omega_x}k_x\big)
\sin^2\big(\frac{\pi}{L\omega_y}k_y\big)}\\
\times\Big[\mu+2-\cos\big(\frac{2\pi}{L\omega_x}k_x\big)-\cos\big(\frac{2\pi}{L\omega_y}k_y\big)\Big]^\frac{1}{2}.
\end{multline}
To extract the large $L$ behaviour of~\eqref{eq:app-p3} it is useful to 
split the integration domains $[0,L\omega_x/2]$ and $[0,L\omega_y/2]$ to write  
\begin{multline}
\langle\vec{1}|\mathbb{P}^{(th)}|\vec{1}\rangle=
\frac{2\sqrt{2}}{\sqrt{g}L^4\omega_x^2\omega_y^2}
\sum_{l_x=0}^{L/2-1}\sum_{l_y=0}^{L/2-1}\int_{l_x\omega_x}^{(l_x+1)
\omega_x}dk_x \int_{l_y\omega_y}^{(l_y+1)\omega_y}dk_y
\\
\times\frac{\sin^2(\pi k_x)\sin^2(\pi k_y)}{\sin^2\big(\frac{\pi}{L\omega_x}k_x\big)\sin^2\big(\frac{\pi}{L\omega_y}
k_y\big)}\Big[\mu+2-\cos\big(\frac{2\pi}{L\omega_x}k_x\big)-
\cos\big(\frac{2\pi}{L\omega_y}k_y\big)\Big]^\frac{1}{2}.
\end{multline}
We now restrict ourselves to the case with $\omega_x=1/p$ and $\omega_y=1/q$, with 
$p,q$ positive integers. After a simple shift of the integration 
variables as $k_x\to k_x-l_x\omega_x$ and $k_y\to k_y-l_y\omega_y$, one obtains  
\begin{multline}
\label{eq:app-p4}
	\langle\vec{1}|\mathbb{P}^{(th)}|\vec{1}\rangle=
	\frac{2\sqrt{2}p^2q^2}{\sqrt{g}L^4}
	\sum_{p'=0}^{p-1}\sum_{q'=0}^{q-1}\sum_{l_x=0}^{L/(2p)-1}
	\sum_{l_y=0}^{L/(2q)-1}\int_{0}^{1/p}dk_x \int_0^{1/q}dk_y\\
	\times\frac{\sin^2(\pi(k_x+l_x+p'/p))
	\sin^2(\pi (k_y+l_y+q'/q))}
	{\sin^2\big(\frac{p\pi}{L}(k_x+l_x+p'/p)\big)
	\sin^2\big(\frac{q\pi}{L}(k_y+l_y+q'/q)\big)}\\
\times\Big[\mu+2-\cos\big(\frac{2 p\pi}{L}(k_x+l_x+p'/p)\big)-
\cos\big(\frac{2q\pi}{L}(k_y+l_y+q'/q)\big)\Big]^\frac{1}{2}.
\end{multline}
We now focus on the behaviour at $g<g_c$. We set 
$\mu=\gamma_4/L^4$ (cf.~\eqref{eq:mu-redef}), and we 
expand~\eqref{eq:app-p4} in the limit $L\to\infty$. This gives 
\begin{multline}
\label{eq:app-p5}
	\langle\vec{1}|\mathbb{P}^{(th)}|\vec{1}\rangle=\\
	\frac{4}{\sqrt{g}\pi^3L}
	\sum_{p'=0}^{p-1}\sum_{q'=0}^{q-1}\sum_{l_x=0}^{L/(2p)-1}
	\sum_{l_y=0}^{L/(2q)-1}\int_{0}^{1/p}dk_x \int_0^{1/q}dk_y
	\frac{\sin^2(\pi(k_x+p'/p))
	\sin^2(\pi (k_y+q'/q))}
	{(k_x+l_x+p'/p)^2(k_y+l_y+q'/q)^2}\\
\times
\Big[\frac{\gamma_4}{2\pi^2 L^2}+ p^2(k_x+l_x+p'/p)^2+
q^2(k_y+l_y+q'/q)^2\Big]^\frac{1}{2}.
\end{multline}
Here we used the periodicity of the trigonometric  functions. 
The term $\gamma_4/L^2$ can be neglected for $L\to\infty$. 
Importantly, as 
a result of the large $L$ limit, Eq.~\eqref{eq:app-p5} 
depends only on the low-energy part of the dispersion of the QSM, 
although it contains non-universal information. 
We now have to determine the asymptotic behaviour of the sum over $l_x,l_y$ 
in~\eqref{eq:app-p5}, i.e., of the function $\eta_{p',q'}(k_x,k_y)$ defined 
as 
\begin{equation}
\label{eq:app-sum}
\eta_{p',q'}(k_x,k_y)=
\frac{4}{\sqrt{g}\pi^3L}
\sum_{l_x=0}^{L/(2p)-1}
	\sum_{l_y=0}^{L/(2q)-1}
\frac{[p^2(k_x+l_x+p'/p)^2+
q^2(k_y+l_y+q'/q)^2]^\frac{1}{2}}{(k_x+l_x+p'/p)^2(k_y+l_y+q'/q)^2}.
\end{equation}
The asymptotic behaviour of $\eta$ in the limit $L\to\infty$ can be 
obtained by using the Euler-Maclaurin formula. Given a function $f(x)$ 
this is stated as 
\begin{equation}
\label{eq:eml}
\sum_{x=x_1}^{x_2} f(x)=\int_{x_1}^{x_2} f(x) dx \\+\frac{f(x_1)+f(x_2)}{2}+\frac{1}{6}\frac{f'(x_2)-f'(x_1)}{2!}+\dots
\end{equation}
In~\eqref{eq:eml} the dots denote terms with higher derivatives of $f(x)$ 
calculated at the integration boundaries $x_1$ and $x_2$.  These can be derived to arbitrary order. 
To proceed, we first isolate the term with either $l_x=0$ or $l_y=0$ in~\eqref{eq:app-sum}. 
The remaining sum after fixing $l_x=0$ or $l_y=0$ can be treated with~\eqref{eq:eml}. 
We define this contribution to the large $L$  behaviour of $\eta_{p',q'}$ 
as $\eta_0$, which is given as  
\begin{equation}
\label{eq:eta0}
\eta_0=\frac{4}{\sqrt{g}\pi^3}\Big[\frac{q}{(k_x+p'/p)^2}
	+\frac{p}{(k_y+q'/q)^2}
	\Big]\frac{\ln(L)}{L}. 
\end{equation}
In the derivation of~\eqref{eq:eta0}  we neglected the boundary 
terms in~\eqref{eq:eml} because they are subleading. 

We are now left with the sums over $l_x\in[1,L/(2p)]$ and $l_y\in[1,L/(2q)]$ in~\eqref{eq:app-sum}. 
These can be calculated again by using~\eqref{eq:eml}. We first apply~\eqref{eq:eml} to the 
sum over $l_x$. We have two contributions. The first one is obtained after 
evaluating the integral in~\eqref{eq:eml} at $x_2=L/(2p)$. After 
expanding the result for $L\to\infty$, we obtain the contribution 
$\eta_1$ given as 
\begin{equation}
	\label{eq:app-ln}
	\eta_{1}=\sum_{l_y=1}^{L/(2q)}\frac{4p}{\sqrt{g}\pi^3(k_y+l_y+q'/q)^2}\frac{\ln(L)}{L}. 
\end{equation}
Note the term $\ln(L)/L$ in~\eqref{eq:app-ln}. 
The sum over $l_y$ in~\eqref{eq:app-ln} can be performed exactly to obtain in the large $L$ limit
\begin{equation}
\label{eq:eta1}
\eta_{1}=\frac{4}{\sqrt{g}\pi^3}p\psi'(1+k_y+q'/q)\frac{\ln(L)}{L}. 
\end{equation}
Here $\psi'(z)$ is the first derivative of the digamma function $\psi(z)$ 
with respect to its argument. 
The second contribution is obtained by evaluating the integral in~\eqref{eq:eml} 
at $x_1=1$. The remaining sum over $l_y$ cannot be evaluated analytically. 
However, one can, again, treat the sum over $l_y$ with~\eqref{eq:eml}. After 
neglecting the boundary terms in~\eqref{eq:eml}, which are subleading for large $L$,  
and after evaluating the integral in~\eqref{eq:eml} at $x_2=L/(2q)$, we obtain 
the contribution $\eta_2$ as 
\begin{equation}
\label{eq:eta2}
\eta_2=\frac{4}{\sqrt{g}\pi^3}\frac{q}{1+k_x+p'/p}\frac{\ln(L)}{L}. 
\end{equation}
Having discussed the contribution which derives from approximating the 
sum over $l_x$ in~\eqref{eq:app-sum} with the integral in~\eqref{eq:eml}, we 
finally focus on effect of the boundary terms in~\eqref{eq:eml}. 
Let us consider the first boundary term (first term in the second 
row in~\eqref{eq:eml}). A term as $\ln(L)/L$ is obtained by 
fixing $l_x=1$, other contributions being subleading. After performing the 
sum over $l_y$ one obtains the first boundary contribution $\eta_{b1}$ as 
\begin{equation}
\label{eq:etab1}
\eta_{b1}=\frac{2}{\sqrt{g}\pi^3}\frac{q}{(1+k_x+p'/p)^2}\frac{\ln(L)}{L}. 
\end{equation}
Similarly, the second boundary term (last term in~\eqref{eq:eml}) 
gives 
\begin{equation}
\label{eq:etab2}
\eta_{b2}=\frac{2}{3\sqrt{g}\pi^3}\frac{q}{(1+k_x+p'/p)^3}\frac{\ln(L)}{L}. 
\end{equation}
Note that boundary terms in~\eqref{eq:eml} are expected to be small. Specifically, 
the $k$-th term is suppressed as $1/(k+1)!$. 
The final result for $\eta(k_x,k_y,p,p',q,q')$ is obtained by putting 
together~\eqref{eq:eta0}\eqref{eq:eta1}~\eqref{eq:eta2}\eqref{eq:etab1}\eqref{eq:etab2} to 
obtain 
\begin{equation}
	\label{eq:group}
	\eta_{p',q'}(k_x,k_y)=\eta_0+\eta_1+\eta_2+\eta_{b1}+\eta_{b2}. 
\end{equation}
%

\subsection{Finite-size contribution}
\label{app:app2}

In this section we derive the leading behaviour in the large $L$ limit of 
$\langle\vec{1}|\mathbb{P}^{(L)}|\vec{1}\rangle$. 
Interestingly, we show that in the presence of a straight boundary between 
the two subsystems (see Fig.~\ref{fig:partition1}) one has the 
behaviour $\langle\vec{1}|\mathbb{P}^{(L)}|\vec{1}\rangle\propto \ln(L)/L$. 
On the other hand, in the presence of corners the multiplicative 
logarithmic correction is absent. 
The finite-size correlator reads as (cf.~\eqref{eq:q2}) 
\begin{multline}
	\label{eq:q2-app}
	\mathbb{P}^{(L)}_{\vec{n}\vec{m}}=-\frac{1}{4\sqrt{g}
	\pi}\sideset{}{'}\sum_{l,l'=-\infty}^\infty
	e^{-\sqrt{2\mu}\sqrt{(l L+n_x-m_x)^2+(l'L+n_y-m_y)^2}}
	\\\times\Big[\frac{1}{[(l L+n_x-m_x)^2+(l' L+n_y-m_y)^2]^{3/2}}
	+\frac{\sqrt{2\mu}}{(l L+n_x-m_x)^2+(l' L+n_y-m_y)^2}\Big].
\end{multline}
Crucially, if $\omega_x<1$ and $\omega_y<1$, the denominators in~\eqref{eq:q2-app} 
are never singular. This implies that the logarithmic correction is not 
present, which can be straightforwardly checked numerically. 
Let us now consider the situation with $\omega_x<1$ and $\omega_y=1$. 
Now, a singularity appears in the limit $L\to\infty$ for $l=0$ and $l'=\pm1$. 
We numerically observe that only the first term in~\eqref{eq:q2-app} gives rise 
to a singular behaviour. Thus, we neglect the second term and fix $l=0$,  obtaining 
\begin{multline}
\label{eq:pp}
	\langle\vec{1}|\mathbb{P}^{(L)}|\vec{1}\rangle=
	-\frac{1}{4\sqrt{g}\pi L^2\omega_x}\,\sideset{}{'}
	\sum_{l'=-\infty}^\infty\sum_{n_x,m_x=0}^{L\omega_x}\sum_{n_y,m_y=0}^{L-1}
	\frac{
	e^{-\sqrt{2\mu}\sqrt{(n_x-m_x)^2+(l' L+n_y-m_y)^2}}}{((n_x-m_x)^2+(l' L+n_y-m_y)^2)^{3/2}}
\end{multline}
Only the differences $n_x-m_x$ and $n_y-m_y$ appear in~\eqref{eq:pp}. Thus, it is convenient to 
change variables to $x=n_x-m_x$ and $y=n_y-m_y$, to obtain 
\begin{multline}
\label{eq:qtilde}
\langle\vec{1}|\mathbb{P}^{(L)}|\vec{1}\rangle=
-\frac{1}{4\sqrt{g}\pi L^2\omega_x}\sideset{}{'}\sum_{l'=-\infty}^\infty\sum_{x=-L\omega_x}^{L\omega_x}\sum_{y=-(L-1)}^{L-1}\\
\Big(L\omega_x+1-|x|\Big)(L-|y|)\frac{e^{-\sqrt{2\mu}\sqrt{x^2+(l' L+y)^2}}}{(x^2+(l' L+y)^2)^{3/2}}. 
\end{multline}
Again, the singular behaviour occurs for $x\approx 0$ and 
$y\approx -l L$, with $l'=\pm1$.  In this limit we can neglect the 
exponential in~\eqref{eq:qtilde} because it is regular. Thus, we obtain 
\begin{equation}
\label{eq:qtilde}
\langle\vec{1}|\mathbb{P}^{(L)}|\vec{1}\rangle=
-\frac{1}{4\sqrt{g}
\pi L^2\omega_x}\,\sideset{}{'}\sum_{l'=-\infty}^\infty\sum_{x=-L\omega_x}^{L\omega_x}\sum_{y=-(L-1)}^{L-1}
\frac{(L\omega_x+1-|x|)(L-|y|)}{(x^2+(l' L+y)^2)^{3/2}}. 
\end{equation}
To proceed, let us now consider the case with $l=1$. It is clear that the contribution 
from $l=-1$ is the same. We can restrict the sum over $x$ in~\eqref{eq:qtilde} to 
$x>0$ because of the symmetry $x\to-x$. We also restrict to $y<0$ because the singularity 
in~\eqref{eq:qtilde} occurs for $y<0$. We now have 
\begin{equation}
\label{eq:qtilde-1}
\langle\vec{1}|\mathbb{P}^{(L)}|\vec{1}\rangle=
\frac{1}{2\sqrt{g}\pi L^2\omega_x}\sum_{x=0}^{L\omega_x}\sum_{y=0}^{L-1}
\frac{(L\omega_x+1-x)(y-L)}{(x^2+(L-y)^2)^{3/2}}. 
\end{equation}
Now the strategy is to treat the sum~\eqref{eq:qtilde-1} by using the Euler-Maclaurin 
formula~\eqref{eq:eml}. For instance, one can first apply~\eqref{eq:eml} to 
the sum over $x$. One obtains that  the leading term in the large $L$ limit 
is obtained by evaluating the integral in~\eqref{eq:eml} 
at $\omega_x L$. One can also verify that the boundary terms in~\eqref{eq:eml} 
can be neglected. 
A straightforward calculation gives 
\begin{equation}
	\label{eq:p-fs-fin}
	\langle\vec{1}|\mathbb{P}^{(L)}|\vec{1}\rangle=-\frac{1}{\sqrt{g}\pi}\frac{\ln(L)}{L}, 
\end{equation}
where the contribution of $l=-1$ in~\eqref{eq:qtilde} has been taken into account. 

\bibliographystyle{SciPost_bibstyle.bst}
\bibliography{bibliography}

\end{document}